
\pageno=0
\footline={\ifnum\pageno=0{}\else\hfil\number\pageno\hfil\fi}
\newif\ifnamedreferences

\namedreferencesfalse

\def\hyperref#1#2#3#4{{#4}}
\def\hyperdef#1#2#3#4{{#4}}

\input amssym.def
\input amssym.tex
\hfuzz5pt
\newcount\equationcount
\newcount\sectioncount
\newcount\subsectioncount
\newcount\subsubsectioncount
\edef\thissection{}
\edef\thissubsection{}
\edef\thissubsubsection{}
\edef\thisequation{}
\edef\empty{}
\def\label#1#2{\edef\temporary{\noexpand\hyperref{}{#1}%
               {\csname this#1\endcsname}{\csname this#1\endcsname}}%
           \ifx\temporary#2\relax
           \else\ifx\undefined#2\relax\global\let#2=\temporary
           \else
             \errhelp{You have used something in a label statement
                      which had an incompatible previous
                      declaration. Are you sure you do not have a bad
                      forward statement before? If you continue, I
                      shall redefine this label!}%
             \errmessage{'\noexpand#2' already defined to be
                         '\expandafter\noexpand#2' not
             '\temporary'}%
             \global\let#2=\temporary
           \fi\fi}
\def\forward#1#2#3{\def#2{\hyperref{}{#1}{#3}{#3}}}
\def\section#1{\global\advance\sectioncount by 1
               \ifnum\sectioncount>0\edef\thissection{\the\sectioncount}\else
                  \let\thissection\empty\fi
	       \let\thissubsection\empty\subsectioncount0
	       \let\thissubsubsection\empty\subsubsectioncount0
               \let\thisequation\empty\equationcount0
               \par\goodbreak\bigskip\hyperdef\loose{section}{\thissection}{%
               \noindent\bf
               \ifx\thissection\empty\else\thissection. \fi
               \uppercase{#1}}\nobreak}
\def\abstract{\begingroup\sectioncount=-1\section{\hfill ABSTRACT\hfill}}
\def\endabstract{\vfil\eject\endgroup}
\def\appendix#1#2{\let\sectioncount0
                \edef\thissection{#2}
	        \let\thissubsection\empty\subsectioncount0
 	        \let\thissubsubsection\empty\subsubsectioncount0
                \let\thisequation\empty\equationcount0
                \par\goodbreak\bigskip\hyperdef\loose{section}%
                {#1.\thissection}{%
                \noindent\bf
                \uppercase{#1 #2}}}
\def\subsection#1{\global\advance\subsectioncount by 1
                  \ifnum\subsectioncount>0\edef\thissubsection{%
                                       \ifx\thissection\empty\else
                                       \thissection.\fi\the\subsectioncount}%
                  \else\let\thissubsection\empty\fi
                  \let\thissubsubsection\empty\subsubsectioncount0
                  \par\goodbreak\medskip\hyperdef\loose{subsection}%
                  {\thissubsection}%
                  {\noindent\bf
                  \ifx\thissubsection\empty\else\thissubsection. \fi
                  {#1}}}
\def\subsubsection#1{\global\advance\subsubsectioncount by 1
                  \ifnum\subsubsectioncount>0\edef\thissubsubsection{%
                                       \ifx\thissubsection\empty\else
                                       \thissubsection.\fi
                                       \the\subsubsectioncount}%
                  \else\let\thissubsubsection\empty\fi
                  \par\goodbreak\medskip\hyperdef\loose{subsubsection}%
                  {\thissubsubsection}{\noindent\sl
                  \ifx\thissubsubsection\empty\else\thissubsubsection.\fi
                  {#1}}}
\def\eqname#1{\global\advance\equationcount by 1
              \edef\thisequation{\ifx\thissection\empty\else
                                 \thissection.\fi\the\equationcount}%
              \label{equation}#1%
              \hyperdef\loose{equation}{\thisequation}{#1}}
\def\eqn#1{\eqno(\eqname#1)}
\def\title#1{{\par \leftskip=0pt plus 1fil\relax\rightskip=\leftskip
              \parindent=0pt \parfillskip=0pt
              \hyphenpenalty=10000 \exhyphenpenalty=10000
              \bf \uppercase{#1}\bigskip}}
\def\defrefs#1#2{\expandafter\def\csname ref.#1\endcsname{#2}}
\newcount\referencecount
\ifnamedreferences
\def\ref#1{{\def\accent{}\if\csname ref.#1\endcsname\relax\iffalse{\else}\fi
            \hyperref{}{reference}{#1}{#1}\else\csname ref.#1\endcsname}\fi}
\def\putref#1#2{\item{\hyperdef\loose{reference}{#1}{[#1]}} #2}
\else
\def\ref#1{{\def\accent{}%
            \expandafter\ifx\csname ref.#1\endcsname\relax
             \global\advance\referencecount by 1
             \expandafter\xdef\csname ref.#1\endcsname
                 {\noexpand\hyperref{}{reference}{\the\referencecount}%
                  {\the\referencecount}}%
            \fi
            \csname ref.#1\endcsname}}%
\def\reftag#1{{\def\accent{}%
            \expandafter\ifx\csname ref.#1\endcsname\relax
             \global\advance\referencecount by 1
             \expandafter\xdef\csname ref.#1\endcsname
                 {\noexpand\hyperref{}{reference}{\the\referencecount}%
                  {\the\referencecount}}%
            \fi }}%
\def\putref#1#2{{\def\accent{}%
                 \expandafter\ifx\csname ref.#1\endcsname\relax
                  \global\advance\referencecount by 1
                   \expandafter\xdef\csname ref.#1\endcsname
                   {\noexpand\hyperref{}{reference}{\the\referencecount}%
                     {\the\referencecount}}%
                 \fi
                 \def\hyperref##1{\hyperdef\loose}
                 \item{[\csname ref.#1\endcsname]} #2}}%
\fi

\def\fig#1{\hyperref{}{figure}{#1}{#1}}
\def\putfig#1{\hyperdef\loose{figure}{#1}{Figure #1}}

\baselineskip=18pt

\title {LARGE  ENERGY  CUMULANTS  IN  THE 2D POTTS  MODEL  AND  THEIR
EFFECTS IN FINITE SIZE ANALYSIS}

\vskip 2. truecm

\centerline {\bf T.~Bhattacharya~$^a$, R.~Lacaze~$^{b,c}$, A.~Morel~$^b$}

\vskip .5 truecm

a. MS B285, Group T-8, Los Alamos National Laboratory, WM 87544, USA

b. Service de Physique Th\'eorique, CEA-Saclay, F-91191 Gif-sur-Yvette Cedex,
France

c. ASCI, Bat. 506, Universit\'e Paris Sud, 91405 Orsay Cedex, France

\vskip 2. truecm

\abstract

We   develop  an ansatz  for  expressing  the free  energy  of the two
dimensional $q$-states Potts model for $q > 4$  near  its first order
phase transition point. We notice that for the moderate  values of $ q
\lesssim  15 $, the    energy profile at   the  phase transition   is not
expressible as a  sum of gaussians.  We discuss  how  this affects the
traditional finite    size analysis of   this   phase transition.   In
particular,  the dominant length   scale   governing the finite   size
corrections turns  out  to be  much  (${} \sim 6$ ~times)  larger than the
largest correlation length in the problem.

\vskip 2. truecm
\rightline{SPhT-94/008}

\endabstract

\section{Introduction}

\label{section}\intro

\reftag{Baxter 73} \reftag{Black 81} \reftag{Wu 82}  \reftag{Borgs 90}
\reftag{Kl\"umper 90} \reftag{Buffenoir 93} \reftag{Borgs 92}

\reftag{Binder 82} \reftag{Peczak 89} \reftag{Lee 90}
\reftag{Gupta 92} \reftag{Billoire 92} \reftag{Berg 92} \reftag{Janke 92}
\reftag{Grossmann 93} \reftag{Billoire 93} \reftag{Billoire 94} \reftag{Janke
94}

There has been a   lot of recent   interest in the numerical  study of
first order transitions. Though these systems  are undeniably of great
phenomenological importance, a purely analytical  study has often been
beyond the reach  of  current methods. The  two-dimensional  $q$-states
Potts model is, in this  arena, a very  useful model: for, on the one
hand, a lot is known  about this model analytically  [\ref{Baxter 73}-
\ref{Buffenoir 93}];
on the other, a number of numerical simulations [\ref{Binder 82}-
\ref{Janke 94}] have been performed on this model.

Unfortunately, the finite size scaling behaviour of the numerical data
are often not  satisfactorily explained by the exact analytical
results at hand.  Small discrepancies also  appear when the
numerically   measured   surface    tensions are compared    to  their
theoretical estimates.

In  an attempt  to  clarify these  issues, we   tried to condense  the
analytically known  results   into  an ansatz which    might correctly
incorporate the  leading singularities   of the free-energy   near the
transition point  as   function  of  $q$ and  $\beta$,    the  inverse
temperature. This allowed us to investigate all  moments of the energy
distribution function at  the  transition point. As expected  from our
previous   analysis     of a  high    $q$   expansion in    this model
[\ref{Bhattacharya 93},\ref{Bhattacharyb 93}] and contrary to popular
expectations, we
discovered that the higher moments are very large at moderate
values of $q \sim 10$. As a  result, the energy distribution cannot be
satisfactorily expressed as  a sum of  gaussian distributions; and the
usual methods of finite size analysis may have to be revised.

In addition to this, the finite volume corrections  appear to follow a
simple scaling  relation,  where, however,  the relevant  length scale
turns out to be {\sl much larger\/}  than the correlation lengths in the
model. In fact only the simulations  performed at a $q$ value as large
as 20 seem to have been in  the truly asymptotic regime [\ref{Billoire 93}].
We believe that these non-negligible  finite size effects (and  proper
accounting of the large energy cumulants) may have caused the apparent
discrepancies in the numerical analyses.

\forward{section}\prelim{2}
\forward{section}\purephase{3}
\forward{section}\ansatz{4}
\forward{section}\freeenergy{5}
\forward{section}\conclusion{6}
The organization of this paper is as follows. In section \prelim, we
summarize the analytically known properties of the model. In the next
section, we show how numerical methods allow to isolate finite volume
effects directly in the free energy. We develop our ansatz in
section \ansatz, and use it to predict a set of scaling relations. We
apply this ansatz to study the available numerical data in section
\freeenergy. We end with a short discussion of the main results of our
paper in section \conclusion.

\section{KNOWN PROPERTIES OF THE MODEL AND PRELIMINARIES}
\label{section}\prelim

The $ q $-states 2-D Potts model is defined by the Hamiltonian
$$ H = - \sum^{ }_{( ij)}\delta_{ \sigma_ i\sigma_ j}\ , \eqn\HEQ $$
where the $ \sigma_ i $ variables, attached to the $ V=L^2 $ sites $ i
$  of  a square lattice,  take $  q $  distinct  values; the summation
extends to all the  $ 2V $  pairs $ (ij) $  of neighbouring sites.  We
shall refer to  $ V $ as  the lattice volume.  Many  properties of this
model have been known  for a long time and  can be found in the review
by Wu[\ref{Wu 82}].

Throughout this paper, we shall call
$$ F = \lim_{V \longrightarrow \infty} \left({1 \over  V} \ \ln
\ Z \right) \eqn\FREE $$
the free energy of the system, where $ Z  $ is its partition function.
The model has a temperature driven phase  transition which occurs at a
value of the inverse temperature $ \beta $ given by
$$ \beta_ t = \ln \left[ \sqrt{ q} +1 \right] \eqn\BETA $$
This value separates the region where the system  is in either one of
the $  q $ possible ordered phases  ($ \beta >\beta_ t $)
from the  disordered  region ($ \beta <\beta_  t  $). When
necessary,  the quantities in an ordered or in
the disordered phase will be labeled by a subscript $  \varphi = o \hbox{
or } d$, respectively.

The transition  is second order for  $ q \leq  4 $, where the critical
properties are   described by  known, $ q    $ dependent,  indices.  In
particular, the specific heat and correlation length exponents $ \alpha
$ and $ \nu $ are the same in all phases and are given by
$$ \eqalignno{ \alpha( q) & = 2(1-2u)/(3(1-u)) & (\eqname\ALPHA) \cr \nu( q) &
= (2-u)/(3(1-u))\ , & (\eqname\NU) \cr} $$
where $ {\rm cos} \left({\pi \over 2} u \right) =  \sqrt{ q}/2 $.
Note in particular that at $q=4$, $\alpha$ and $\nu$ coincide:
$$ \alpha( 4) = \nu( 4) = 2/3. \eqn\ALPHANU $$
For  each phase  $   \varphi $, we  observe   that although  the
correlation length $ \xi_ \varphi $ and  the specific heat $ C_\varphi
$ both diverge for $ q \leq 4 $  when $ \beta -\beta_ t \longrightarrow
0 $ as
$$ \eqalign{ \xi_ \varphi  & \propto \lambda_ \varphi \left\vert \beta
-\beta_  t \right\vert^{ -\alpha(  q)}+...   \cr C_\varphi &   \propto
\mu_\varphi   \left\vert    \beta  -\beta_   t  \right\vert^{    -\nu(
q)}+... \cr} \eqn\CRIT $$
their ratio remains finite at $ q=4 $ in the neighbourhood of $ \beta_
t.  $   In the  expression above, the   dots  stand for (unknown) less
singular and regular terms at $ \beta =\beta_ t  $.

The phase transition at $ \beta  =\beta_ t $  is first order for $ q>4
$.    The internal energies of  the  pure phases,   $ E_\varphi $, are
exactly known at $ \beta =\beta_ t $ [\ref{Baxter 73}]. The latent heat $
{\cal L} = E_d-E_o $, finite for any $ q>4 $, vanishes  as $ q \longrightarrow
4_+ $; so that this quantity is continuous at the (second order) point
$  q=4 $. Various  expressions and properties of  the $ E_\varphi $'s
 are given  in Appendix A  for  completeness and  further
reference.

In  this first  order  region, the  specific heats  $  C_\varphi $ and
correlation lengths $ \xi_ \varphi $ are finite at any value of $\beta$. The
difference $ C_d-C_o $ is known at $ \beta_ t$:
$$ C_d-C_o = \beta^ 2_t\ {\cal L}/ \sqrt{ q}\ , \eqn\CDIFF $$
which shows  that it  vanishes  in the same way  as   $ {\cal L}  $ when  $ q
\longrightarrow 4_+$.
However, their  sum is unknown.  Exact expressions
for various  correlation lengths   have  been obtained  in  the recent
years [\ref{Kl\"umper 90},\ref{Buffenoir 93}] and details  are
collected in Appendix A.  Here  we quote only  those results which are
relevant to us. According to these  calculations, and to their
interpretation  [\ref{Buffenoir 93},\ref{Borgs 92}]  in
terms  of  the  pure phase $  \xi_   \varphi $'s, the disordered  phase
correlation length $ \xi_ d $ has the following divergent behaviour as
$ q \longrightarrow 4_+$:
$$ \xi_ d \simeq  {1 \over 8 \sqrt{ 2}} x\ , \eqn\XID $$
where $x$ is given by
$$ \eqalignno{  x & = {\rm  exp} \left[\pi^  2/(2\theta) \right] & \cr
\theta &   = \ln \left[{  \sqrt{ q}+   \sqrt{  q-4} \over 2}
\right]\ . & (\eqname\xwz) \cr} $$
The behaviour (\XID) was anticipated by Black and Emery a long time ago
using renormalization group techniques [\ref{Black 81}].
Only very recently numerical simulations were able to produce data in
agreement with this result [\ref{Janke 94}].

When $ q  \longrightarrow  4_+   $, $  2\theta   $ tends  to 0  as   $
(q-4)^{1/2} $ so that  $ x  $  diverges quite strongly.  On  the other
hand,  as  $ q $  increases  away from  4,  $ \theta  $ increases very
slowly, growing only logarithmically as $  q \longrightarrow \infty $,
so that  $  x $ stays  very  large  over a very  broad  range  of  $ q
$-values.  As  a consequence,  the approximation (\XID)  of $  \xi_ d $
remains valid    over    an equally large   domain
[\ref{Borgs 92}] (see  Appendix    A  for  the  size  of  the
corrections to (\XID)). For this reason, the variable $  x $ happens to
be relevant to our forthcoming analysis  of the phase transition above
$ q=4 $. We note in particular that  it also controls the leading behaviour of
the  latent heat, which vanishes  as $ x^{-1/2} $ at  $ q=4  $.

So, the correlation  length at $  \beta =\beta_ t  $ in the disordered
phase becomes infinite as $  q \longrightarrow 4_+ $  as it does at  $
q=4 $ when $ \beta \longrightarrow \beta_{ t_-} $. Less is known about
$ \xi_ o $, but it has been  conjectured by Borgs and Janke
[\ref{Borgs 92}] that
$$ \xi_ o=\xi_ d/2 \eqn\XIO $$
which is consistent with numerical estimates of $\xi_o$ at $ q=10  $
[\ref{Peczak 89},\ref{Gupta 92}].

If $ \xi_ \varphi $ diverges at  $ q \longrightarrow 4_+  $, it is not
unexpected that the specific heat does so as well, since the latter is
related to the integral of the energy-energy correlation function over
the whole lattice:
$$ C_\varphi  = \lim_{V \longrightarrow \infty}  {\beta^ 2 \over 4} \sum^{ }_{
\ell} \left( \left\langle E_{\ell_ 0}E_{\ell} \right\rangle_ \varphi -
\left\langle E_{\ell_ 0} \right\rangle_ \varphi \left\langle E_{\ell}
\right\rangle_ \varphi \right)\ . \eqn\CPHI $$
Here $ \ell_ 0 $ is a fixed link, $ E_{\ell} $ (0 or  1) is the energy
carried by the link $ \ell $, $ \langle \rangle_ \varphi $ denotes the
average inside the phase $ \varphi $; and  the thermodynamical limit $
V  \longrightarrow   \infty $   is taken before   the   limit  $ \beta
\longrightarrow \beta_ t $. As  it generically does  at  a $ 2^{  {\rm
nd}} $  order point, the  sum  in (\CPHI) most probably  diverges along
with the correlation length, the   major contribution coming from  the
region of large $ \ell_ 0-\ell $ separation. Since $ \xi_ d $ is known
to diverge as $ q \longrightarrow 4_+ $,  $ C_d $ should also diverge.
Further, since  $ C_d-C_o $ is  bound to vanish in  the same limit, $ C_o $
also must go to  infinity, consistent with the Borgs-Janke  conjecture
(Eq.~\XIO).

We    continue this review  of  known  properties by  writing down the
consequence of duality for the free energy $ F(\beta)  $ of this model.
At $ \beta \geq  \beta_ t $ (respectively $\beta \leq
\beta_ t   $), $   F(\beta) $  coincides   with  the  ordered
(disordered)  free energy  $  F_o $  ($ F_d $).  Duality
relates $ F(\beta) $ to $ F \left(\tilde \beta \right) $, where
$\beta$ is related to $\tilde\beta$ by
$$  \left( {\rm e}^\beta -1  \right) \left(  {\rm e}^{\tilde \beta} -1
\right) = q\ , \eqn\DUALB $$
and we have
$$ F_{\rm d}  \left(\tilde   \beta \right)-  \ln  \left[{ {\rm  exp}
\left(\tilde \beta \right)-1  \over \sqrt{ q}}   \right] = F_{\rm o}(\beta)  -
\ln \left[{ {\rm exp}(\beta)   -1 \over \sqrt{  q}} \right]\
. \eqn\DUALF $$
It is convenient to  parametrize the distance to $  \beta =\beta_ t  $
(the transition point, where  $ \beta =\tilde \beta  $) by  defining a
variable $ b $ through
$$ {\rm e}^b = { {\rm e}^\beta -1 \over \sqrt{ q}}\ . \eqn\B $$
As $ \beta \longrightarrow \beta_ t $, $  b $ vanishes as $ \left(1+1/
\sqrt{ q} \right)  \left(\beta -\beta_ t  \right) $, the interchange $
\beta  \longleftrightarrow\tilde     \beta $      is   just     $    b
\longleftrightarrow -b $,  and duality is  the statement that the free
energies of the model can be written
$$ \eqalignno{  &  F_o(\beta)  = F_t+b+g(b) &  \cr \noalign{\vbox to
0pt{\vss\vbox{\hbox to \hsize{\hfil (\eqname\DUALOD)}}\vss}}
  &  F_d(\beta)  = F_t+b+g(-b)\ . &  \cr} $$
Here  $ F_t $  is   $ F_o \left(\beta_   t \right)=F_d  \left(\beta_ t
\right) $, $ b $ is defined by (\B), and  $ g(b) $, which vanishes at
$ b=0 $, contains all the essential information about $ F(\beta) $. It
follows the $ n^{\hbox{th}} $ energy cumulants at $
\beta =\beta_ t $
$$ F^{(n)}_\varphi \equiv  \left. { {\rm d}^n \over  {\rm d} \beta^ n}
F_\varphi( \beta) \right\vert_{ \beta =\beta_ t}\ , \eqn\FN $$
once expressed in terms of the derivatives $ g^{(p)}(0) $, $ p\leq n $, of $
g(b) $ have the same
form for $ \varphi =o $ and $ d $, up to the change $ g^{(p)}(0)
\longleftrightarrow  (-)^pg^{(p)}(0) $.
Thus, the
combination $ F^{(n)}_o - (-)^nF^{(n)}_d $ is completely determined by
the knowledge of the  cumulants of order lower than   $ n. $  Equation
(\CDIFF) and the  relation
$$ E_d+E_o =  -2 \left(1+1 \sqrt{ q} \right) \eqn\ESUM $$
are two well known examples of this general statement of duality.

Up   to now  we   only  considered properties  of   the model   in the
thermodynamical limit; in  particular  the limit  $ V  \longrightarrow
\infty  $ was always taken  before  the limit  $ \beta \longrightarrow
\beta_ t $, which  allowed us to properly define  $ F(\beta) $. We now
turn  to the properties  of the system  in  a finite box with periodic
boundary   conditions,    and    recall  the     result  obtained   in
Ref. [\ref{Borgs 90}] for the   corresponding finite
volume partition
function $ Z_V(\beta) $:
$$ Z_V(\beta) = q\  {\rm exp} \left[VF_o(\beta)   \right] + {\rm  exp}
\left[VF_d(\beta) \right] + R(L,\beta) \ . \eqn\ZVB $$
The first two terms  represent the $ q   $ ordered and  the disordered
phase contributions to the total  partition function respectively, and
the essential statement is that the rest $ R(L,\beta) $ is bounded by
the inequality
$$  \left\vert  R(L,\beta) \right\vert <  c_1  \times q^{-c_2 L}
\times \mathop{\rm exp}(VF(\beta)) , \eqn\REST $$
where $c_1$ and $c_2 > 0$ are constants. Similar statements hold  for the
logarithmic derivatives  of $ Z_V(\beta) $, that  is for the cumulants
of the energy distribution.  We refer to $R(L,\beta)$ as a ``true''
finite size effect, and to the first two terms in Eq.~(\ZVB) as
the asymptotic ordered and disordered contributions.

We also recall that
$$ \eqalign{ F(\beta) & \equiv F_o(\beta) \ \ \ \ {\rm  for} \ \ \ \ \
\beta \geq \beta_ t \cr F(\beta) & \equiv F_d(\beta) \ \ \ \ {\rm for}
\ \ \ \ \ \beta \leq \beta_ t\ , \cr} $$
so  that $   F = \mathop{\rm  Max}  \left(F_o,F_d  \right) $. Strictly
speaking, the result  holds for $  q $  \lq\lq large enough\rq\rq.  We
will ignore this caveat, and consider that (\REST) is valid  for $ q >4
$.

Eqs.~(\ZVB,\REST) are very interesting since they control the size of any
deviation of the partition function  at finite size from the  familiar
two component representation $ Z = qZ_o+Z_d $. For this reason, we are
going to  demonstrate in the next section how $ F(\beta) -F_t $, close to
$\beta
= \beta_t$ can, in principle,
be extracted from very precise numerical data on
the energy distribution,    supplemented  by  the    duality  property
(\DUALOD).

\vskip 17pt

\section{NON ASYMPTOTIC PURE PHASE FREE ENERGIES FROM NUMERICAL DATA}
\label{section}\purephase

By definition, the partition function is given by
$$ Z_V(\beta) =   \sum^{ }_ E\Omega_  V(E) {\rm  exp}[-\beta VE] \eqn
\PART $$
where $  VE $ is the energy  of a configuration,  and $ \Omega_ V(E) $
(which is independent of $ \beta  $), the number of configurations with
that energy in  the  given volume. The  current  numerical simulations
measure  the energy probability distribution at  some  value of
 $  \beta  $, say $ \beta_ 0 $:
$$  P_{V,\beta_  0}(E)  = \Omega_ V(E)  {\rm  exp}  \left[-\beta_  0VE
\right]/Z_V \left(\beta_ 0 \right)\ . \eqn\PVB $$
It follows from the two above equations that
$$  {Z_V(\beta) \over  Z_V  \left(\beta_  0  \right)}  = \sum^{ }_   E
P_{V,\beta_ 0}(E) {\rm e}^{-VE \left(\beta -\beta_ 0 \right)}\ , \eqn
\ZVBBO $$
which  yields the $  \beta  $ dependence of $   Z_V(\beta)  $ up  to a
constant factor $ Z_V \left(\beta_ 0 \right) $, and thus, an effective
free energy via
$$ F^{ {\rm eff}}(L ,\beta) -  F^{ {\rm eff}} \left(L, \beta_ 0\right)
=   {1 \over  V} \ln \left[Z_V(\beta) /Z_V \left(\beta_  0
\right) \right]\ . \eqn\ZBBO $$
This is true for any  lattice system, but,  in such generality, is not
very useful  for finite size studies because  the $  V $ dependence of
the constant  ({\it i.e.\hbox{}}  $  \beta $-independent)  $  F^{ {\rm
eff}} \left(L,\beta_ 0 \right) $ remains unknown.

It becomes more interesting in the case of a  system which has a first
order transition at $ \beta =\beta_ t $, with, as in the Potts model,
a duality property relating the $  q $  ordered phases to  the disordered
phase coexisting  around $
\beta_ t $ in a finite volume.
Consider  the ``asymptotic'' form    $ Z_{
\hbox{as}}(\beta) $  given by (\ZVB) when  the remainder $ R(L,\beta) $
is  dropped and   $  F_o $, $  F_d   $  are  represented in   the form
(\DUALOD). Then we construct the quantity
$$ X(b) \equiv {Z_{ {\rm  as}}(\beta) \over Z_{ {\rm as}} \left(\beta_
t \right)} {\rm exp}(-bV) =  {q\ {\rm exp}[Vg(b)] + {\rm  exp}[Vg(-b)]
\over q+1} \eqn\XB $$
with $ b  $ given  by (\B)  as a function  of  $ \beta $.  As explained
above,  the  function $  X(b)    $ can   be measured  from   numerical
simulations in the vicinity of $ b=0  $ (that is of  $ \beta =\beta_ t
$). The function $ g(b) $ then follows from the knowledge of  $ X $ at
$ b $ and $ -b $:
$$ g(b) = {1 \over V} \ln \left[{qX(b) - X(-b) \over q-1} \right]\ ,
\eqn\GEXP $$
and corresponding $ F_o^{\rm eff} $  and $ F_d^{\rm eff}$ can  be
reconstructed from Eq.~\DUALOD.  By
construction, $ g(b) $ should be independent of the lattice
size  in the limit  where the  remainder $  R(L,\beta)  $, bounded  by
(\REST), can be neglected.  In turn, any  $V$ dependence is a
 {\sl direct  measure\/}  of ``true'' finite  size effects,   and these
effects,
as opposed to the situation of the  energy   distribution  (see  our
discussion  in the   next sections),  are  under  theoretical  control
[\ref{Borgs 90}].

Let us now  show that this  extraction of  finite size  effects in the
pure phase free energies is indeed feasible. For  this purpose, we use
the   numerical data of  Ref.~[\ref{Billoire 92}], for
the $ q=10  $ Potts model with linear  lattice sizes $  L = 16$,  20,
24,   36,  44 and  50.   Details on the  runs  can  be   found in the
reference. We just recall that a  multi-spin coding was used, allowing
for  the simultaneous    generation of  16   independent Monte   Carlo
histories and that at each lattice size, a very long run was made near
the maximum of the specific heat. We use the data from these long
runs and all the errors quoted  below are computed from the observed
dispersion  about their mean of the  16 corresponding results. Because
it turns out that the finite size effects are  very small, a plot of $
g(b) $ itself as given  by Eqs.(\GEXP, \XB{} and \ZVBBO)  does not
show any $ L $ dependence. For clarity, we choose to show, as a
function of $ \beta $, and for the various lattice sizes, the quantity
$$ W_o[L,\beta] = F_o^{\rm eff} (L,\beta) -F_t+E_o \left(\beta -\beta_ t
\right)\
, \eqn\WO $$
which differs from the  measured effective free energy  $ F_o(L,\beta)
-F_t   $ by the  known  asymptotic linear term.   Then the finite size
effects   clearly show   up,   as exhibited   in  Figs.~(\fig{1a}) and
(\fig{1b}).   On  the largest  lattices ($  L=50   \hbox{ and  } 44$),
because  of important   statistical errors,  the quantity  $  W_o $ is
available in  a limited range  of $ \beta $  only.  Below $ \beta
=\beta_ t $, this  is due to the  large cancellations occurring in the
argument of the log  in Eq.(\GEXP).  At  large positive values of $ V
\left(\beta -\beta_ t \right) $,  the errors become large  again since
we have to extrapolate data  far away from  the $  \beta $ values where
they   were measured [\ref{Billoire 92}].    Fig.~(\fig{1a})  shows $
W_o(L,\beta) $ for $L=50\hbox{,  }44\hbox{ and }36$. From this figure,
we conclude that, within the present errors, there is no sizable $ L $
dependence for $ L \gtrsim 36 $, in the accessible $ \beta $ range,
as was observed in Ref.~[\ref{Billoire 92}] for
$\left\langle E \right\rangle_{\beta_t}$.

On  the   contrary, finite size  effects    are clearly  exposed  (see
Fig.~\fig{1b}) on lattices of linear  size $ L  < 36 $. In particular,
the slope of $ W_o(L,\beta)   $ at $  \beta  =\beta_ t  $ is not  zero
showing that the ordered phase internal energy is not $  E_o $ (but it
does tend  rapidly to $  E_o $ as  $ L $  increases). This finite size
effect  should {\sl not\/}  be   confused with that  observed for  the
location of the peak in the energy distribution: as will be emphasized
later on,  a  peak displacement  {\sl   must be  present even  in  the
asymptotic regime\/} ({\it  i.e.\hbox{}}  even when $ R(L,\beta)  $ is
negligible). A  closer look  at Fig.~\fig{1b}  also reveals  that  the
curvature  (specific heat) at   $ \beta_ t  $ (as  well  as the higher
derivatives with respect to $ \beta $) do exhibit $ L $ dependence.

The precision reached in the simulation used here does not allow
for a quantitative study of the $ \beta $ and  $ L $ dependence of the
``non  asymptotic'' part,  $  R(L,\beta) $, of  the partition
function, at least in the absence of some model for these effects. What
is demonstrated is the  feasibility of measuring the non-asymptotic
contributions  to the  Potts model  free  energies. We now  turn to a
theoretical  study of  the  asymptotic ordered   phase free  energy  $
F_o(\beta)  $, which  the    above numerical determination  could   be
compared with.

\section{A THEORETICAL ANSATZ WHICH ACCURATELY DESCRIBES THE POTTS
MODEL FREE ENERGY}
\label{section}\ansatz

We start  from the observation made  in Section~\prelim{} that the correlation
length of the model tends to infinity when the point $q=4,
\beta=\beta_t(4)$ is reached along either of the following two paths
in the $q,\beta$ plane:

\vskip 10pt

(i) $ q=4 $,\nobreak\ \nobreak\  $ \beta \longrightarrow \beta_ t(4) $
 (Temperature driven $2^{ {\rm nd}} $ order transition)

(ii)  $ \beta =\beta_  t(q) $, $ q \longrightarrow 4_+  $ (along the $
1^{ {\rm st}} $ order line).

\vskip 10pt

In  case (i) we know  furthermore  that in each   of the two phases $
\varphi =(o,d) $  the ratio of  the
specific   heat  to  the correlation   length   tends  to  a  constant
(Eq.(\CRIT)).

For  case  (ii), we argued  that since  $ \xi_  \varphi $ diverges, it
looks reasonable to believe that $ C_\varphi $ also diverges.

We now conjecture that  in fact $ C_\varphi  /\xi_ \varphi $ is finite
also along path (ii), and formulate our  ansatz as a generalization
of this kind of continuity of $ C_\varphi /\xi_  \varphi $ at $ q=4 $,
$  \beta  =\beta_ t(4)  $.  (To be specific,   we discuss the
ordered  phase $ \left(\beta \longrightarrow   \beta_{ t_+} \right) $,
using  duality  afterwards to   determine  the  free
energy in the other phase.)

In situation (i) the known specific heat exponent implies that
the most singular part  of the second derivative of the
free energy with respect to $ \beta $ is
$$ F^{(2)}_o(\beta) \sim A \left(\beta -\beta_  t \right)^{-2/3}\ , \eqn
\FOSB $$
so that the corresponding $ (p+2)^{\hbox{th}} $ derivative is
$$ F^{(p+2)}_o(\beta) \sim A(-)^p {\Gamma( p+2/3) \over \Gamma( 2/3)}
\left(\beta -\beta_ t \right)^{-2/3-p}\ .\eqn\PPSDER $$
On the other hand, we use our knowledge of the correlation length
exponent:
$$ \xi_ o \sim \lambda_ o \left(\beta -\beta_ t \right)^{-2/3}\ , \eqn
\CORREXP$$
to eliminate $ \left(\beta -\beta_  t \right) $
and obtain the relation
$$  F_o^{(p+2)}(\beta)  \sim A(-)^p  {\Gamma(   p+2/3) \over \Gamma(  2/3)}
\left(\xi_ o/\lambda_ o \right)^{{3p \over 2}+1}\ , \eqn\FXI $$
between the order $ (p+2) $ cumulant of the energy distribution and the
correlation length.

Our ansatz consists in boldly continuing  Eq.(\FXI), from situation (i)
to  situation (ii),  where   we know the    behaviour  of $  \xi_  o $
(Eqs.(\XIO,\XID)) as a function  of $ q $ at  $ \beta =\beta_ t $. Hence
we have
$$ F^{(p+2)}_o  \left(\beta_ t \right) \sim  A(-)^p {\Gamma( p+2/3) \over
\Gamma( 2/3)} (Bx)^{1+3p/2}\ , \eqn\FPT $$
where  $ x $ is  defined in Eq.(\xwz). The  constant $ B $ includes the
proportionality constants  occurring in $ \xi_ o  $ as a function of $
\beta $ or of $ x $. Let us add a few comments.

\vskip 10pt

a) All these relations express the
leading behaviour  as either $ \beta \longrightarrow  \beta_  t $ at $
q=4  $, or  $ q \longrightarrow  4_+   $ at $   \beta  =\beta_ t  $. In
particular,  subdominant terms in  the  basic equation (\FPT) arise
from less singular or regular terms in both $ F^{(2)}_o $ and $ \xi_ o $ at
the second order point $  q=4 $, $  \beta =\beta_ t(4) $, neglected above.
Including the (known)  corrections to the low
$ (q-4) $ behaviour of $ \xi_ o $ alone would thus have been inconsistent.

b) We recall that $ x $ is very large over a wide range of values of $ q
> 4   $.
Accordingly,   all   the  energy cumulants   $   F^{(p+2)}_o
\left(\beta_ t  \right)  $,   which behave   like  $
x^{(1+3p/2)} $, become  extremely large if $p  \geq 0$.  If we  set $
p=-1  $,  Eq.(\FPT)   yields    a  vanishing  contribution  of   $   O
\left(x^{-1/2} \right) $ to the  internal energy $  E_o $. Though this
must  be   neglected  in  comparison with   other  contributions to  $
F^{(1)}_o(\beta)   $ which do    not vanish  at $     q=4 $, $   \beta
\longrightarrow \beta_ t(4) $, we note that the known result
$$ E_o=- \left(1+1/ \sqrt{ q} \right)-{\cal L}/2, \eqn\EXACTEO $$
shows the presence of this term as the latent heat $ {\cal L} $ vanishes as $
x^{-1/2} $ (see Appendix) in the limit $ q \longrightarrow 4_+ $.

c)   At the leading order   in  $ x  $  we  consider here, the  energy
cumulants  $ F^{(p+2)}_d $  in the disordered  phase  are given by the
expression (\FPT) with  the $ (-)^p  $ sign dropped. This follows  from
duality (see Eq.(\DUALOD)) which implies
$$ F^{(p+2)}_d  = (-)^p F^{(p+2)}_o +  O \left(F^{(p+1)}_o  \right)\ ,
\eqn\DTOO $$
where  the corrections are down by $ x^{3/2}  $ compared to the leading
term.

d) In Eq.~(\FPT), we  call   the critical  amplitude,  $  A $,  and  the
proportionality factor,   $  B $,  \lq\lq  constants\rq\rq .  What  we
actually mean is that they are smooth functions of $ q $ when compared with
the violent behaviour of $ x(q) $. This is indeed  what appears in the
latent heat $  {\cal L} $ (Appendix), which is  $ x^{-1/2} $ times a function
of $ q $  which varies smoothly from  $ \pi $ at  $ q=4 $  to $ 4\pi /
\ln \ q $ as $ q \longrightarrow \infty  $.

e) For consistency with duality of our ansatz (\FPT),  we have the same
$ A $ and $ B $ in the ordered and disordered phases cumulants. That $
A_o=A_d $ is no  surprise  since duality  applied directly to  (\FOSB)
(which is valid at $ \left(\beta -\beta_ t \right)\longrightarrow 0+ $)
yields
$$ F^{(2)}_d(\beta) \ \sim \  A \left(\beta_ t-\beta \right)^{-2/3}, $$
which in turn is valid for $ \left(\beta_  t-\beta
\right)\longrightarrow 0+ $. That  it is so also
for the factor $ B $ may seem more intriguing since $ \xi_ o $ is used
in deriving (\FPT) and is different  from $ \xi_  d $. What it
actually implies is that the  ratio $  \xi_  d/\xi_ o $  at $   q>4 $,
which is  two according to (\XIO), must also be two  at $
q=4 $ ({\it i.e.\hbox{}} it must be continuous at $ q=4_+ $).

\vskip 10pt

With the expression (\FPT) of the  derivative $ F^{(n)}_o(\beta) $ at $
\beta_  t $,  $  n\geq 2   $, and the    knowledge of the   first order
derivative $ -E_o $, we  obtain for $
F_o(\beta) - F_o \left(\beta_ t \right) $:
$$ F_o(\beta) - F_o \left(\beta_  t \right) = -E_o \left(\beta -\beta_
t \right)  + {3A \over(  Bx)^2} \left[{3 \over 4} \left(1+ \left(\beta
-\beta_ t \right)(Bx)^{3/2}   \right)^{4/3}-{3 \over 4} -  \left(\beta
-\beta_ t \right)(Bx)^{3/2} \right]\ . \eqn\FDIFF$$

For later convenience in exploiting this expression, we rearrange it and
rescale its variables. After setting
$$ \eqalignno{ S^2 & = (Bx)^2/(3A) &  \cr v &  = \left(\beta -\beta_ t
\right)(Bx)^{3/2} &\cr\noalign{\hbox to\hsize{\hfill\vbox to
0pt{\vss\hbox{(\eqname\SCALE)}\vss}}} \varepsilon_  o & = E_o(Bx)^{1/2}/(3A)
& \cr \varepsilon_ d & = E_d(Bx)^{1/2}/(3A)\ , & \cr} $$
and noting that,  to the order of  accuracy retained, $  F_d $ follows
from $ F_o $ by changing $  v $ into  $ -v $  and $ E_o  $ into $ -E_d
$,(\footnote{$ ^\ast     $}{there       is   a   misprint        in
Ref.~\ref{Bhattacharya 93} where $ \varepsilon_  d $ should be
changed into $ -\varepsilon_ d $ in Eq.(12).}) we obtain the essential
result of this section
$$ \eqalignno{ S^2 \left(F_o(\beta) -F_t \right) & =  - {3 \over 4} -v
\left(\varepsilon_  o+1 \right) + {3   \over 4}(1+v)^{4/3} &
(\eqname\ANSATZ o)
\cr   S^2  \left(F_d(\beta)  -F_t  \right) &    =  - {3  \over   4} -v
\left(\varepsilon_ d-1 \right)  +  {3 \over 4}(1-v)^{4/3}  &
(\ANSATZ d)
\cr} $$

These are the expressions we propose as ans\"atze for the free energies
of the model  in the  first  order  region. According  to  the way  we
established them,  we expect their   domain of validity  to  be the
region where $ Bx $ is
large and  $ v $ is finite. Under  these  conditions, we immediately note
that

\vskip 10pt

(i) $ S $ (which is proportional to  the correlation length) plays the
role of a length scale, to be compared with the linear scale  $ L $ of
the finite lattice  on which we compute  the partition function.

(ii) After rescaling  the internal energy parameters  $ E_o $, $ E_d $
according  to  (\SCALE)  the  \lq\lq asymptotic\rq\rq{}
partition function
$$ Z   =    q\ {\rm  exp}  \left(L^2F_o(\beta)   \right)  +  {\rm exp}
\left(L^2F_d(\beta) \right) $$
depends on $ L $ and $ q $  {\it only} through  the reduced size $ L/S
$ (apart for the multiplicity $q$).
This statement is true up to residual dependencies of $ A $ and $ B
$ upon $ q $ in (\SCALE) (see comment (d) above).

\vskip 10pt

We repeat that the expressions (\ANSATZ o,\ANSATZ d) are just ans\"atze
which  have been  built from the {\sl assumed\/} continuity  of  ratios of
energy cumulants at $  q=4 $ and above, a  continuity which we have no
theoretical argument  for.  Hence, we have  to justify our  approach by
explicit  comparison of its  consequences   with results of  numerical
simulations.   This will  be done  in the  next section.  However,  we
already want to point out  that Eqs.~(\ANSATZ o and \ANSATZ d) suffer  from
the absence of singularity at  $ \beta =\beta_ t $  (that is $ v=0 $).
Such a singularity is  expected [\ref{Fisher 67}] or proven  to
exist  [\ref{Isakov 84}] in the  similar  situation of a  field
driven first order transition at  low temperature (Ising model below $
T_c $ in  the limit of a vanishing  magnetic field). Here  we take the
heuristic  point  of view   that this  singularity,  although probably
essential, is {\it  numerically} mild, so that  it does not affect the
practical consequences we shall now discuss. A probably related weakness
of our ansatz is its singularity at $v=-1$ for $F_o$ ( and $v=+1$ for
$F_d$). Indeed $(1+v)^{4/3}$ is the only analytical function at $v=0$
whis has the required $F_o^{(n)}(\beta_t)$. But an essential singularity
at $v=0$, with a vanishing contribution to all derivatives, may alter the
analyticity structure at $v<0$.

\section{APPLICATIONS OF OUR ANSATZ FOR $ F(\beta) $}
\label{section}\freeenergy

In order to exploit  the consequences of the  ansatz (\ANSATZ), we need
to assign numerical values to the \lq\lq constants\rq\rq{} $  A $ and $
B  $  at the value of  $  q $  we are  interested in. As  explained in
Ref.~[\ref{Bhattacharya 93}], we are  able to do so by comparing
the values of the second and third derivatives of  $ F_o(\beta) $ at $
\beta =\beta_ t $, given by
$$ \eqalign{  F^{(2)}_o \left(\beta_ t \right)  & = ABx,  \cr F^{(3)}_o
\left(\beta_ t \right) & = - {2 \over 3} A(Bx)^{5/2}\ , \cr}\eqn\SECTHIRD $$
to the analytical results of a  large $ q $  expansion. A full account
of  the  procedure,  which includes  suitable  Pad\'e  resummations of
series in $ 1/ \sqrt{ q} $ now extended to order  10, is given in
Ref.~[\ref{Bhattacharya 94}], together  with estimates of $ A
$ and $ B $ for the $ q $ values of interest.

\subsection{Comparison of $ F_o(\beta) $ with numerical data at $
q=10 $}

As exhibited  in  Section  \purephase,  a  simulation  measuring the    energy
distribution  in a   region of  $  \beta  $ close to   $  \beta_  t $,
supplemented by duality, gives access to  an estimate,
$ F_o^{\rm eff}(L,\beta) $,
of  the ordered  free  energy. This  estimate  should  coincide with $
F_o(\beta) $  in the asymptotic limit defined   by the criterion that
$ R(L,\beta) $ can be neglected in  Eq.(\ZVB). In Fig.~(\fig{2}), we
thus show the difference
$$ \Delta_ o \equiv F_o^{\rm eff}(L,\beta) - F^{( {\rm ansatz})}_o(\beta) \
. \eqn\DEL $$
The quantity $ F_o^{\rm eff}(L,\beta) $ is the $ F_o $ extracted up to
its   constant    term    from  the   $     q=10 $      simulation  of
Ref.~[\ref{Billoire 92}] using Eqs.~(\DUALOD, \GEXP, \XB, and \ZVBBO)
while $ F^{ {\rm ansatz}}_o $ is the prediction of Eq.~(\ANSATZ o), where
we used the predetermined values [\ref{Bhattacharya 93}]
$$ \eqalignno{ A & = .193 & \cr B & = .386\ \ \ \ \ \ \ \ {\rm at} \ \
\ \ q=10 & (\eqname\AB) \cr} $$
We did not plot the data for $ L=44 \hbox{ and } 50$, which are always
compatible  with  those  for  $  L=36   $,  with  larger   errors (see
Fig.~\fig{1a}). Let us now comment on Fig.~\fig{2}.

The lower limit of the $ \beta $ range shown  corresponds to the value
$ -1 $ of the scaled variable $ v $ of Eq.~(\SCALE), a value below which
our ansatz becomes  meaningless. The upper limit corresponds  to $ v $
about 2 and  $ \beta $ values above  which the data at $  L \geq  36 $
become unreliable. The $ \beta $ range can also  be expressed in terms
of the variable $ \left(\beta -\beta_ t \right)V $, a natural argument
for the partition function: it extends from about -4  to about 8 for $
L=36 $. In this interval where we can compare the data and the ansatz,
the following observations hold:

\vskip 10pt

(i) Although clearly visible, the finite size effects remain tiny:
the maximum  absolute  value   of  $  L^2  \Delta_ o$
is less than $0.02$ for all the data shown.

(ii) For $ \left\vert v\right\vert < 1 $, the difference between the $
L=36 $ data and the ansatz is less than 1 standard deviation.

(iii)  As already mentioned,  a quantitative  study  of  the $ (L,\beta)  $
dependence is difficult to achieve. The  slope at $  \beta =\beta_ t $
is,  however, measurable  and shows  a departure from the asymptotic
value   $ -E_o=1.66425  $.  We   find  respectively at  $  L=16\hbox{,
}20\hbox{, }24\hbox{ and }36 $
$$  10^3\times  \left(E_o(L)-E_o    \right) =   -9.0(8),\  -5.1(1.0),\
-1.2(1.0), \hbox{ and } 1.6(3.0)\ . \eqn\LIST$$
This clearly indicates a   \lq\lq fast\rq\rq{} decrease of  the  finite
size  effect, of  the  same order of  magnitude  as the  one found  in
[\ref{Billoire 92}] for the total average energy at $ \beta
=\beta_ t $, and compatible with a  behaviour $ {\rm exp} \left(-L/L_o
\right)  $ with $  L_o \sim 7$.
Note that this value is
intermediate between $  \xi_ d=10.6 $  which should be relevant to the
description of    {\it    interface effects}   in   $    R(L,\beta)  $
[\ref{Borgs 92}], and $ \xi_  o=\xi_ d/2 = 5.3 $ expected
to control the finite size effect in the pure ordered phase itself.

We conclude that  our ansatz does  provide an accurate description  of
the   asymptotic free energies in   a substantial range   of $ \beta $
values around $ \beta =\beta_ t $. It can be used as a reference model
for a quantitative study  of  finite size  effects measured in  a high
precision numerical  study  for which  the   moderate size  range  $ L
\lesssim 40 $ would be especially suited.

\subsection{Energy distributions and finite size effects}

Given the  free energies  $   F_o(\beta) $    and  $ F_d(\beta)   $,  the
corresponding energy distribution  can be computed by  inverse Laplace
transform  of Eq.~(\PART). As  before, we  call \lq\lq asymptotic\rq\rq{}
the distribution  obtained when $  R(L,\beta) $ is neglected in (\ZVB),
that is when
$$ Z_V(\beta)   \simeq  Z^{ {\rm  as}}_V(\beta) \equiv    q\ {\rm exp}
\left[V\  F_o(\beta) \right] +  {\rm exp} \left[V\ F_d(\beta) \right]\
. \eqn\ASYMPDEF$$
The probability distribution at $ \beta =\beta_ o $ is then given by
$$ P^{ {\rm as}}_{V,\beta_ o}(E) =  N \int^{\bar \beta +i\infty}_{\bar
\beta -i\infty} {\rm d} \beta  \ {\rm exp} \left[V \left(\beta -\beta_
o \right)E \right]Z^{ {\rm as}}_V(\beta) . \eqn\PVE $$
In this formula, $ N $ is a normalization factor fixed  for each $ V $
and $ \beta_ o $ by
$$ \sum^{ }_ EP_{V,\beta_ o}(E) = 1\ , \eqn\NORMALCOND$$
$ \bar \beta $ is suitably chosen  according to analyticity properties
of  $ Z(\beta) $. Note  that  strictly speaking, $ F_o  $  and $ F_d $
should be periodic functions of complex $ \beta $  with period $ 4i\pi
V $ since  the allowed  values  of $ E  $  in Eq.~(\PART) on a  finite
lattice   are of  the   form $   E=-k/(2V)   $,  $ k=0,1,\ldots,2V  $.
Correspondingly, the integration contour in  (\PVE) should extend over
a finite range, {\it e.g.\/\hbox{}}, from  $ \bar \beta -2i\pi  V $ to $ \bar
\beta  +2i\pi V $. We  neglect such refinements:  any reasonable $ Z^{
{\rm as}} $ used in computing $ P(E) $ should not depend on details of
$ F(\beta) $ so far away from the real axis.  This will be the case in
what follows.

Equation~(\PVE) of course   implies that quadratic approximations  of $
F_o $ and $  F_d $ as  functions of $ \left(\beta  -\beta_ t \right) $
lead to   the   celebrated two  gaussian   formula  [\ref{Binder 86}]
for the probability distribution.  Higher terms in $ F_o $
and $ F_d   $ generate size   dependent deformations of the two   peak
structure observed in the   vicinity  of the first order   transition,
which should not be misinterpreted as  `non-asymptotic' contributions in
the    partition      function.  This     caveat     does not     hold
[\ref{Borgs 90}] if the energy cumulants  at $
\beta =\beta_  t $ only are analyzed.

Because, as we  argued  in previous  sections, the  Potts model  free
energies receive very  large  non quadratic contributions, $   F^{(n)}
\left(\beta_ t \right) \sim x^{{3n \over 2}-2} $ for  a broad range of
$  q $ values, this  model  offers a valuable  laboratory for studying
their effect in  measured probability distributions.  With this remark
in mind we will now compare to numerical data the predictions of (\PVE)
when $ F_o $, $ F_d $ are given by our ansatz (\ANSATZ o,\ANSATZ d).

\vskip 10pt

\subsubsection{Comparison with numerical data}

For definiteness,  we choose   $  \beta_ o=\beta_ t   $ in  (\PVE). The
integral to be computed is thus:
$$ \everymath{\displaystyle} \matrix{  P_{V,\beta_ t}(E)= N \int^{\bar
v+i\infty}_{\bar v-i\infty} {\rm d} v \times {\rm exp} \left[ \left({L
\over S} \right)^2\varepsilon v  \right] \times \hfill \cr  \hfill \cr
\times  \left\{   q\ {\rm exp}  \left\{   \left({L  \over S} \right)^2
\left[-v \left(\varepsilon_ o+1 \right)+{3 \over 4}(1+v)^{4/3} \right]
\right\}  + {\rm  exp} \left\{  \left({L \over   S} \right)^2 \left[-v
\left(+\varepsilon_   d-1   \right)+{3   \over 4}(1-v)^{4/3}   \right]
\right\} \right\} \hfill \cr} \eqn\EDIS $$
Here we use  the variables $ S  $,  $ v $,  $ \varepsilon_  o  $ and $
\varepsilon_ d $ defined in Eq.~(\SCALE) and also scale the energy $ E $
according to
$$ \varepsilon  = E(Bx)^{1/2}/(3A)\ . \eqn\E $$
$ N $ is a new normalization factor. This form (\EDIS) shows explicitly
that, apart from the multiplicity $q$ of the ordered phase contribution,
 $ P(E) $ depends on $ q $ and  $ L $ through $  (L/S) $ alone when
expressed as a function of $ \varepsilon $ variables.

At large $ (L/S)^2 $ one may certainly evaluate the integral (\EDIS) by a
saddle
point method. The question we are trying to address here is how  large
should $ L  $ be for this to be true.  Taking $ q=10 $  for
definiteness,    using the numerical     values  from (\AB)  and  the
expressions (\SCALE) and (\xwz), we find
$$ S_{q=10} = 60.3\ , \eqn\STEN $$
about six times larger than the largest correlation length $ \xi_ d $!
Since we want to compare with distributions measured at $ L \leq 50 $,
it   is clear that  the  saddle point method   is completely
inadequate for this
purpose. We shall use it later only for obtaining simple expressions on
effectively large  lattices. For the  time being we  have to perform a
numerical integration of (\EDIS).

Up to trivial changes, contributions in the disordered and ordered phase
need similar analyses: only the ordered one is considered here. We have to
choose $ \bar v $ and a convenient path to $ \pm  i\infty $.  We first
remark that the stationary point of the integrand is
$$ v_S(E) = -1 + \left(\varepsilon_  o-\varepsilon +1 \right)^3, \eqn
\SP$$
a value for which the integrand is well defined for
$$ \varepsilon_ o-\varepsilon +1 \geq  0\ .\eqn\EPSILONRANGE $$
In  such cases, we  take $ \bar  v  = v_S(E) $  and  the path $ v=\bar
v+i\lambda , $ $ \lambda \in[ -\infty ,+\infty] $. The term $ {3 \over
4}(1+v)^{4/3}  $ insures an exponential  fall off of  the integrand at
large $ \left\vert \lambda\right\vert $.

For the other case, $  \varepsilon \geq \varepsilon_ o+1  $, we fix  $
\bar v $ at $ -1 $ and again choose $ v-\bar v $ purely imaginary.

Our results for $ L=16, $ 20, 24, 36, 44  and 50 ({\sl with no
arbitrary adjustable parameters\/}) are shown on Figs.~(\fig{3a},\fig{3b})
as     continuous     curves,  together     with     the   data     of
[\ref{Billoire 92}]. (Note  the  logarithmic scale.)  The
overall agreement, which continues over nearly three decades for
$P(E)$, is very good,
it becomes nearly excellent at the largest $ L $ values. In particular
{\sl the non-gaussian character of the peak is very well reproduced}.

We already   know from our  study   of  Section~\purephase{} that  non
asymptotic contributions  are present when $ L  < 36 $.   We see their
manifestation in Fig.~(\fig{3a}). Since the curves have unit area, the
discrepancies between theory and   numerical data are concentrated  in
the largest (the ordered) peak  region, and are emphasized by plotting
$ P(E) $ in  that region on  a non-logarithmic  scale (Fig.~\fig{4a}):
the  predicted peak looks too narrow.   However, this does {\sl not\/}
mean that, at such $ L $  values, the effective ordered specific heat,
or $   F^{(2)}_o  \left(L,\beta_ t   \right)  $, is   larger than  its
asymptotic value: in fact   Fig.~\fig{2}  shows that the converse   is
true. At  least for $ L=16 $  and 20, the  effective broadening of the
peak  arises  from   large  asymmetries  $   ( \left\vert    F^{(3)}_o
\left(L,\beta_ t \right)  \right\vert $ larger  at low $ L $  values),
whereas $ F^{(2)}_o $ is actually smaller  (the curves in Fig.~\fig{2}
are concave around $\beta_t$). Hence, when seen in energy distributions,
the nature of finite size effects can easily be misidentified. At the
lower right corner of Fig.~(4a), our curves level up; this unphysical
feature is due to the tail of the ordered (left) peak. It disappears
exponentially in $L^2$ and is due to the singularity at $v=-1$ in
(\ANSATZ o) (see below the evolution of the ordered contribution far
away from the peak).

At  larger $  L $    values (Fig.~\fig{3b}),  the agreement is   considerably
improved;  in particular the absolute prediction  for the ordered peak
at $ L=50 $ is  perfect (Fig.~\fig{4b}).  However, a  new effect tends
to  become
sizable: a small  discrepancy appears in  the dip region between the
two peaks at $ L=50 $. This is the region  where interface effects are
expected to emerge  [\ref{Binder 82}],  eventually yielding a
flat plateau in $ E $  at $ \beta =\beta_ t  $. We shall soon discuss
the effective $ L $ dependence of the dip depth, of direct physical
interest for interface tension determinations.

In order  to understand better what is  significant and what is not in
finite  size effects   observed in the   energy  distributions, we now
examine what the expected  shape of $ P_V(E) $  is when $ L  $ becomes
actually large compared with the scale $S$ (${} \sim 6 \ \xi_ d $ at $
q=10 $) introduced in Section~\ansatz.

\subsubsection{Effective $ L  $ dependence and large  $ (L/S) $ limit of
energy distributions}

Amongst  the quantities  which have  been  extracted, and  tentatively
interpreted, from numerically determined energy distributions, are the
peak locations $ E^{ {\rm max}}_{o,d}(L)  $ (asymptotically $ E_{o,d})
$ and widths (asymptotically related to $ C_{o,d}) $, and the quantity
$$ 2\sigma^{ {\rm  eff}}_{od}  = {1 \over L}  \ln \left[{P_V
\left(E^{  {\rm max}} \right) \over P_V  \left(E^{ {\rm dip}} \right)}
\right]\ , \eqn\SOD $$
(asymptotically   twice      the ordered-disordered  interface tension
[\ref{Binder 82}]). For definiteness, we discuss these quantities at $
\beta =\beta_ t $.

Let us  start with a study  of the integral  (\EDIS) via a saddle point
method at large $ \ell =L/S $.  Since by duality one peak is trivially
deduced from the knowledge of the other, we again focus on the ordered
peak, the first  contribution in the integrand  of (\EDIS).
The saddle point $v_S(E)$ of (\SP)
is useful when it lies above the singular point of the ansatz, $
v=-1 $, that is in the energy region
$$ R1\ :\ \ \ \ -2 \leq  E < E_o + 3A/(Bx)^{1/2}. \eqn\ESPO $$

$  R_1 $ extends  above $ E_o  $ by a  small fraction (about 12\% at $
q=10  $) of  the latent  heat.  It is the  situation  of interest  for
evaluating    the   asymptotics    of  the peak      structure.  It is
straightforward to find the saddle  point result for the ordered phase
energy distribution inside $ R_1 $. We find,  up to an energy
independent factor, that
$$ P^{ {\rm ordered}}_V(E\in  R1) = \left[1+\varepsilon_ o-\varepsilon
\right] \ {\rm exp} \left\{ -\ell^ 2 \left[{ \left(1+\varepsilon_
o-\varepsilon  \right)^4 \over 4} - \left(1+\varepsilon_ o-\varepsilon
\right) \right] \right\} \left(1+O \left(1/\ell^  2 \right) \right)\ ,
\eqn\PSPO $$
$$\ell =L/S $$
where  we   recall that $   \varepsilon  -\varepsilon_  o= \left(E-E_o
\right)(Bx)^{1/2}/3A $.   Note  that $   P $  is
asymptotically   the exponential of  a   polynomial of  degree  4. The
prefactor corrects it  by a term $  1/\ell^ 2 $  smaller.  In order to
analyze  the  approach to  asymptotics  of this  peak,  we expand  the
quantity
$$ p_o(\varepsilon) \equiv -{1 \over \ell^ 2}  \ \ln  \ P^{ {\rm
ordered}}_V(E) \eqn\POEQN$$
around its maximum $  \tilde \varepsilon_ o  $, consistently up to $O(1)$
in $ 1/\ell^ 2 $. The result is
$$   \tilde \varepsilon_ o  = \varepsilon_  o-1/  \left(3\ell^ 2 \right)
\eqn\EMAXSP $$
and
$$ \eqalign{ p_o(\varepsilon) & = {\rm  cst.} + {3 \over 2} \left(1+{1
\over  \ell^   2}  \right)  \left(\varepsilon  -\tilde  \varepsilon_ o
\right)^2 - \left(\varepsilon - \tilde \varepsilon_ o  \right)^3 \cr & +
{1 \over 4}   \left(1+{1   \over \ell^ 2}   \right)  \left(\varepsilon
- \tilde \varepsilon_ o   \right)^4   +   O \left({1 \over   \ell^   2}
\left(\varepsilon  -\tilde \varepsilon_  o  \right)^5 \right)\  . \cr}
\eqn\po $$
Corrections of order $ 1/\ell^ 2 $ are generic to the saddle point
method.  The  above  expansion is specific   to our ansatz   (note the
absence of   $ 1/\ell^ 2 $  term  in the cubic term,   and of any term
surviving at $ L=\infty $ at order $ \geq 5 $ in $ \varepsilon -
\tilde \varepsilon_ o  $).   It predicts the  approach to  asymptotics of the
ordered phase peak location, of the specific heat (quadratic term) and
of the asymmetry (cubic  term), and is relevant  to a comparison  with
recent numerical simulations at $ q=10 $  and 20 where such quantities
have been considered [\ref{Billoire 94}]. Since $  \ell $ is not large
in most  cases we also compare the   latter results with those  of the
numerical integration of (\EDIS), from which we compute for each $ q $
and $ L   $  the peak location  and   curvature at  the  peak.   These
comparisons  are  shown  in  Figures~\fig{5a},\fig{5b},\fig{5c}    and
\fig{5d} for $ E^o_{  {\rm  max}} $ and Figures~\fig{6a},\fig{6b}  and
\fig{6c} for $ C_o $.

Fig.~\fig{5a} shows $ E^o_{ {\rm max}} $ for $ q=10 $ as a function of
$ 1/L $ (instead of the natural $ 1/L^2)  $ in order to emphasize once
for all how misleading the finite size effects in energy distributions
can be.  Indeed, the data points of [\ref{Billoire 94}] as well as the
continuous curve (our ansatz)  exhibit an essentially linear behaviour
in $ 1/L $ over the whole range $ L=12\mathop{\hbox{--}}100 $.  In the
absence  of the  known exact  value (square symbol)   at $ 1/L=0 $,  a
linear fit to  the data points would be  reasonable, but would predict
an asymptotic value wrong by many standard deviations!  This is not so
for the $ q=20 $ case (Fig.~\fig{5b}),  which reflects the fact that $
L/S $ is  indeed large  for  this case.   Hence the  apparent  $ 1/L $
behaviour clearly is a  finite size artefact, and  should not be taken
as a  significant property    of the distributions    [\ref{Lee  90}].
Figs.~\fig{5c} and \fig{5d} show the same quantities plotted against $
1/L^2  $,  together with  the first   order correction (straight line)
predicted by Eq.(\EMAXSP).    The adequacy of this representation   is
clearly  exposed, as well as   the strong deviations from  asymptotics
observed in  the data and  well (although  not perfectly) explained by
our ansatz.

We choose to present  $ C_o $ (as  extracted from the curvature at the
peak  maximum) as a function of  $ S^2/L^2 $,  in Figs.~(\fig{6a}) and
(\fig{6b}) for $ q=10 $ and  20 respectively.  Similar conclusions can
be drawn. Note the ranges of $ S^2/L^2 $  in the graphs are much larger
than the domain of  applicability of the
saddle  point method: $  S^2/L^2 \ll  1$!  As an  illustration of  the
scaling property associated with Eq.~(\EDIS), we  finally show, on the
same  plot for  $ q=10  $  and  20,  the ratio $  C_o(L)/C_o(\infty) $
against $ S^2/L^2  $ (Fig.~\fig{6c}). We see  that the data points fall close
to  a same curve  and to the theoretical   prediction, even in regions
where the asymptotic expectation is very far away.

Let us now turn to the energy  region in between  the two peaks, which
according to [\ref{Binder 82}] can be (and has been) used for
a determination   of  the   interface  tension   between  ordered  and
disordered phases. Let us recall the argument. At $  \beta =\beta_ t $
and  for  sufficiently large  linear  size  $ L $,   the  mixture of a
disordered phase in a fraction $ \alpha V $ of the volume separated by
two walls from an ordered phase  occupying the fraction $ (1-\alpha) V
$ has a relative weight which is

\vskip 10pt

i) independent of $ \alpha $ because $ \alpha F_o+(1-\alpha) F_d = F_t
$ at $ \beta =\beta_ t $.

ii) proportional   to  $ {\rm  exp} \left[-2\sigma_{  od}L   \right] $
because each wall costs a free energy $ \sigma L $.

\vskip 10pt

\noindent Thus, a plateau in energy  is expected, for any  $ E $ of the
form
$$ E = \alpha E_o +(1-\alpha) E_d, \eqn\ELATE$$
whose height with  respect to the rest  of the distribution is $  {\rm
exp} \left[-2\sigma_{ od}L  \right]  $.

The predicted  plateau has indeed been  seen  in recent simulations of
the Potts model,  either  by  extending the  linear  size of  a  {\sl
square\/} lattice  to many times  the  correlation length  ($ q=20 $  in
[\ref{Billoire 94}]),  or by  using rectangular
lattices, which favour the  appearance of separation walls  transverse
to   the    large   dimension  ($   q=7     $  in  [\ref{Grossmann 93},
\ref{Gupta 92}]). Such  simulations  have led  to  determinations of  $
\sigma_{  od} $ in very good  agreement with  the predicted $ \sigma_{
od}=1/ \left(2\xi_ d \right) $ of Ref.~[\ref{Borgs 92}].

It   is quite remarkable  that previous   numerical attempts on square
lattices with sizes $ L  \leq 100 $, not  large enough at $ q \lesssim
10 $ for the plateau to develop, already led to satisfactory (although
not as good) estimates of $ \sigma_{ od} $ (Ref.~[\ref{Berg 92}] for $
q=10 $  and Ref.~[\ref{Janke 92}]  for  $ q=7 $).    Indeed, as we saw
above  (Fig.~\fig{3a},\fig{3b}) our ansatz   is able to  reproduce the
data  for $ q=10  $, $  L \lesssim 50   $ quite correctly, although it
totally  ignores contributions  from  mixed phase  configurations.  We
illustrate this fact in Fig.~\fig{7a} where we show our result and the
data of [\ref{Billoire 94}] at $ q=10 $ for the quantity
$$ 2f_{od}   = {1 \over L}   \ln   \left\{{  \sqrt{ P^{ {\rm
max}}_oP^{ {\rm max}}_d} \over P_{ {\rm min}}} \right\} \eqn\twof $$
as a function  of $ 1/L  $. The curve  is computed  from the numerical
integration leading  to  the  energy distributions already   shown. It
levels out at low values of $  1/L $ (dotted  part). This is because
in the region
$$ R2 :\ \ \ \ \ \ E_o + {3A \over( Bx)^{1/2}} < E < E_d - {3A \over(
Bx)^{1/2}}\ , \eqn\RTWOREGION$$
the leading contribution to $ P_{o,d} $ as  $ L/S $ becomes large, as
obtained by a saddle point method, is proportional to
$$ {\rm exp} \left[-L^2  \left\vert E-E_{o,d}  \right\vert /(Bx)^{3/2}
\right]\ , \eqn\RTWOCONTRIB $$
so that the right hand  side of (\twof) grows as  $ L $. Only in  this
regime mixed phase contributions, of relative weight proportional to $
{\rm exp} \left(-2\sigma_{ od}L  \right) $ may  show  up, and this  is
supported by the data. What is puzzling is that for $ L \lesssim 40 $,
our  curve  accounts   for  the trend    of   the data;  moreover   an
extrapolation to $ 1/L =0 $ of its nearly  linear part would accidentally
produce a reasonable result. The same remark applies to the case $ q=7
$, illustrated in Fig.~\fig{7b} (There the definition of $ f_{od} $ is
slightly different: we follow the definition of [\ref{Janke 92}]). Our
conclusion is  that it may  be misleading  to identify $  f_{od} $, as
defined by (\SOD) or (\twof) or as in [\ref{Janke 92}],
to the interface tension $ \sigma_{ od} $ whenever a clear plateau
is not seen in between the two peaks.

\section{Conclusions}
\label{section}\conclusion

The main result of this paper can be summarized as that the ansatz for
the free energy developed in section \ansatz{} is in qualitative and
quantitative agreement with available numerical data. Our work
indicates that, at least for the Potts model at moderate values of
$q$, the energy distribution near a first order transition point cannot be
accurately described as a sum of two gaussian contributions. This
statement is distinct from the correct statement that, up to
exponentially small (what we call non-asymptotic)
corrections, the partition function {\sl can\/} indeed be written as a sum of
the partition function of the pure phases.

These large energy cumulants  of  the     {\sl asymptotic\/}
distributions call for more careful analysis of the numerical data. In
particular, the finite size  effects do not  seem to be  controlled by
the  correlation lengths of the  model: rather,  the relevant scale is
significantly larger. This    implies that numerical  work  should not
assume  that any  system is  sufficiently large: one can ({\it
e.g.\hbox{}} using Eq.\hbox{} \GEXP) and must specifically look for ``true''
finite size effects in the system (in the sense of Eq.~(\ZVB) ).
Especially, one must be careful about
interpreting the depth of the minimum of the energy distribution as a
measurement
of the surface tension: true surface tension effects lead to a flat
minimum which does appear on a sufficiently large lattice.
In  addition, deviations from predictions
of the two-gaussian model (e.g. discrepancies in the position and curvature
of the peaks in the energy distribution),  should  not, in itself  be
considered a finite  volume effect: non-quadratic terms in the free
energy expansion do lead to such shifts.

The non-gaussian nature of the {\sl pure
phase\/} energy distributions near the transition point might seem unusual
and pathological to the particular model at hand. In reality,
our ansatz indicates that this is probably due to the
influence of the nearby second order transition starting at $q=4$. One
may conjecture that strong deviations from the gaussian nature are not
unusual in systems without a large surface tension between the phases.
Such might also be the case of phenomenologically significant models
like QCD, and the results of this paper might be relevant to any
numerical simulation of these phase transitions.

We thank J.M. Luck for his careful reading of our manuscript and for
useful comments.

\vfill\eject

\appendix{appendix}{A}

For the $ q $-states 2D Potts model on a  square lattice, the internal
energies $ E_o   $ and $ E_d  $  in the ordered and  disordered phases
respectively are given for $ q > 4 $ by [\ref{Baxter 73}]
$$ \eqalignno{ E_d+E_o & = -2 \left(1+1/ \sqrt{ q} \right) & (\eqname\one) \cr
{\cal L}\equiv E_d-E_o & = 2 \left(1+1/  \sqrt{ q} \right) {\rm tanh} {\theta
\over 2} P(\theta) & (\eqname\two) \cr} $$
with
$$ 2\ {\rm cosh} \ \theta = \sqrt{ q} \eqn\three $$
and
$$ P(\theta) =  \prod^{ \infty}_{n= 1}( {\rm tanh}  \ n \theta)^  2\ . \eqn
\four $$

Two expressions of $ P(\theta) $ [\ref{Baxter 73}] are of great
interest  for accurate numerical  estimates of  the  latent  heat $  {\cal L}
$. One is very rapidly converging at large $ q $:
$$  \sqrt{  P(\theta)}  =   1  + 2 \sum^{   \infty}_{  m=1}(-)^m  {\rm
e}^{-2m^2\theta} \ , \eqn\five $$
the  other one  is extremely accurate  at small  $ \theta  $
$$ \sqrt{ P(\theta)} = \left({2\pi  \over \theta} \right)^{1/2} \sum^{
\infty}_{  m=0}   {\rm   e}^{-   \left(m+{1   \over 2}   \right)^2\pi^
2/(2\theta)} \eqn\six $$
In the  paper, we chose to  discuss all quantities in  the low $ q-4 $
regime as functions of the variable $ x $ defined as
$$ x = {\rm exp} \left[-\pi^ 2/2\theta \right]\ , \eqn\seven $$
so that in this regime we have
$$ {\cal L}  =  h(q) x^{-1/2} \left[1+2x^{-2}+O \left(x^{-4}  \right) \right]
\eqn\eight $$
with
$$ h(q) = 2 \left(1+1/  \sqrt{ q} \right) {\rm  tanh} {\theta \over 2}
\times \left({2\pi \over \theta} \right) \eqn\nine $$
a smooth  function compared to $  x $. The  corrections to the leading
term  in (\eight) are extremely small,  still less than 3\% for  $ q $ as
large as 100. From these properties, we see that there is a very large
domain of $ q $  values where large $ q  $ as well as  small $ (q-4) $
expansions (\five) or (\six) yield very accurate estimates.

Duality relates the specific heat difference to $ {\cal L} $:
$$ C_d-C_o  = \beta^ 2_t {{\cal L} \over  \sqrt{ q}} \simeq  O \left(x^{-1/2}
\right)\ . \eqn\ten $$
Since we argue that $ C_d $ and $ C_o $  are both $ \sim x
$ at leading order, setting $ C_d=C_o $ as implied by the final ansatz
implies a relative  error  of order $ x^{-3/2},  $  not under  control
within our approach.

The interest of the $ x $ variable also appears in expressions for the
correlation      lengths      [ \ref{Kl\"umper 90}, \ref{Buffenoir 93},
\ref{Borgs 92}].

According to  the above references,   the disordered phase correlation
length $  \xi_ d $ may be  accurately computed either  at  large $ q $
using
$$  1/\xi_  d   =  4 \sum^{   \infty}_{ m=1}    \left[{  \left(-  {\rm
e}^{-\theta} \right)^m \over  m}  {\rm sinh} \left({m\theta  \over  2}
\right)   {\rm tanh}(m\theta) + 2\  {\rm    log} \left({ {\rm cosh}  \
3\theta /4 \over  {\rm  cosh} \ \theta /4}   \right) \right]\ ,  \eqn
\eleven $$
or at low $ (q-4) $ using
$$ \eqalign{  1/\xi_  d & =   4  \sum^{ \infty}_{  m=0} \ln
\left({1+w_m \over 1-w_n}  \right) \cr w_m & =  \left( \sqrt{ 2}\ {\rm
cosh} \left[ \left(m+{1    \over 2}  \right)  \pi^  2/\theta   \right]
\right)^{-1}. \cr} \eqn\twelve $$
In particular the latter expression yields
$$ \xi_ d =  {1  \over 8 \sqrt{ 2}}  x  \left[1-{2x^{-2} \over 3} +  O
\left(x^{-4} \right) \right]\ , \eqn\thirteen $$
subject to comments similar to those made on Eq.~(\eight). The accuracy of
(\thirteen) was already emphasized in Ref.~[\ref{Borgs 92}].

\vfill\eject

\centerline{{\bf REFERENCES}}

\vglue 1truecm
\leftskip=1.5truecm
\let\stop.
\catcode`\.=\active
\def.{\stop\spacefactor=1000\relax}
\putref{Baxter 73}R.J. Baxter, {\sl J. Phys.\/} {\bf C6} (1973)
L-445; {\sl J. Stat. Phys.\/} {\bf 9} (1973) 145.

\putref{Black 81}J.L. Black and V.J. Emery, {\sl Phys. Rev.\/}
{\bf B23} (1981) 429.

\putref{Wu 82}F.Y. Wu, {\sl Rev. Mod. Phys.\/} {\bf 54} (1982)
235.

\putref{Borgs 90}C. Borgs and R. Koteck\'y, {\sl J.
Stat. Phys.\/} {\bf 61} (1990) 79; C. Borgs, R.
Koteck\'y and S. Miracle-Sole,
{\sl J. Stat. Phys.\/} {\bf 62} (1991) 529.

\putref{Kl\"umper 90}A. Kl\"umper, A. Schadschneider and J.
Zittarz, {\sl Z. Phys.\/} {\bf B76} (1989) 247; A. Kl\"umper,
 {\sl Int. Journal of Mod. Phys.\/} {\bf B4} (1990) 871.

\putref{Buffenoir 93}E. Buffenoir and S. Wallon, {\sl J.
Phys.\/} {\bf A26} (1993) 3045.

\putref{Borgs 92}C. Borgs and W. Janke, {\sl J. Phys.\/}
(France) {\bf I2} (1992) 2011.

\putref{Binder 82}K. Binder, {\sl Phys. Rev.\/} {\bf A25}
(1982) 1699.

\putref{Peczak 89}P. Peczak and D.P. Landau, {\sl Phys. Rev.\/}
{\bf B39} (1989) 11932.

\putref{Lee 90}J. Lee and J.M. Kosterlitz, {\sl Phys.
Rev. Lett. \/}{\bf 65} (1990) 137; {\sl Phys. Rev.\/} {\bf B43} (1991) 3265;
J. Lee, {\sl Phys. Rev. Lett.\/} {\bf 71} (1993) 211.

\putref{Gupta 92}S. Gupta and A. Irb\"ack, {\sl Phys. Lett.\/}
{\bf B286} (1992) 112.

\putref{Billoire 92}A. Billoire, R. Lacaze and A. Morel,
{\sl Nucl. Phys.\/} {\bf B370} (1992) 773.

\putref{Berg 92}B.A. Berg and T. Neuhaus, {\sl Phys. Rev.
Letters\/} {\bf 68} (1992) 9.

\putref{Janke 92}W. Janke, B.A. Berg and M. Katoot,
{\sl Nucl. Phys.\/} {\bf B382} (1992) 649.

\putref{Grossmann 93}B. Grossmann and S. Gupta, \lq\lq
Interface Tensions and Perfect Wetting in the Two-Dimensional
Seven-State Potts Model\rq\rq, Preprint HLRZ 63/93.

\putref{Billoire 93}A. Billoire, T. Neuhaus and  B.A. Berg
{\sl Nucl. Phys. B\/} 396 (1993) 779.

\putref{Billoire 94}A. Billoire, B.A. Berg and T.
Neuhaus, {\sl Nucl. Phys. B\/} (FS) 413 (1994) 795.

\putref{Janke 94}W.~Janke and S.~Kappler,
{\sl  Nucl. Phys. B\/} (Proc. Suppl.) {\bf 34} (1994) 674.

\putref{Bhattacharya 93}T. Bhattacharya, R. Lacaze and A. Morel,
{\sl Europhys. Lett.\/} {\bf 23} (1993) 547.

\putref{Bhattacharyb 93}T. Bhattacharya, R. Lacaze and A. Morel,
{\sl  Nucl. Phys. B\/} (Proc. Suppl.) {\bf 34} (1994) 671.

\putref{Fisher 67}M.E. Fisher, {\sl Physics\/} (N.Y.) {\bf 3}
(1967) 255.

\putref{Isakov 84}S.N. Isakov, {\sl Commun. Math. Phys.\/} {\bf
95} (1984) 427.

\putref{Bhattacharya 94}T. Bhattacharya, R. Lacaze and A.  Morel,
\lq\lq Large q expansion of the 2D q-states Potts model\rq\rq,
Preprint Spht-94/066.

\putref{Binder 86}K. Binder, M.S. Challa and D.P. Landau,
{\sl Phys. Rev.\/} {\bf B34} (1986) 1841.

\vfil\eject

(\putfig{1a}) Graph of $W_o[L,\beta]$ (Eq.~\WO) at various large
values of L, in an attempt to look for finite size effects at
$q=10$. Data from Ref.~[\ref{Billoire 92}].

(\putfig{1b}) Graph of $W_o[L,\beta]$ (Eq.~\WO) for small lattices
at $q=10$. Finite size effects clearly show up.
Data from Ref.~[\ref{Billoire 92}].

(\putfig{2}) Graph of $\Delta_o$ (Eq.~\DEL), the difference of the
measured free energy and the free energy predicted by our Ansatz.
Data from Ref.~[\ref{Billoire 92}].

(\putfig{3a}) Comparison of the energy distribution measured on small
lattices against our prediction. Data from Ref.~[\ref{Billoire 92}].

(\putfig{3b}) Comparison of the energy distribution measured on large
lattices against our prediction. Data from Ref.~[\ref{Billoire 92}].

(\putfig{4a}) Enlarged view of the peak region of \fig{3a} showing the
difference between measured and predicted energy distributions.

(\putfig{4b}) Enlarged view of the peak region of \fig{3b} showing the
difference between measured and predicted energy distributions.

(\putfig{5a}) Plot of the peak of the energy distribution versus $1/L$
at $q=10$. A linear fit seems good, but predicts the wrong
extrapolated value. Data from Ref.~[\ref{Billoire 94}].

(\putfig{5b}) Plot of the peak of the energy distribution versus $1/L$
at $q=20$. Data from Ref.~[\ref{Billoire 94}].

(\putfig{5c}) Plot of the peak of the energy distribution versus $1/L^2$
at $q=10$. Data from Ref.~[\ref{Billoire 94}].

(\putfig{5d}) Plot of the peak of the energy distribution versus $1/L^2$
at $q=20$. Data from Ref.~[\ref{Billoire 94}].

(\putfig{6a}) Plot  of the   specific heat  in  the ordered  phase  at
transition for $q=10$ against the  scaled lattice size, compared to our
prediction. Here $C_o$ is extracted from the peak curvature.
Data from Ref.~[\ref{Billoire 94}].

(\putfig{6b})   Plot of the specific   heat   in the ordered phase  at
transition for $q=20$ against the scaled lattice size, compared to our
prediction. Here $C_o$ is extracted from the peak curvature.
Data from Ref.~[\ref{Billoire 94}].

(\putfig{6c}) Plot of the the scaled specific heat at transition
compared to our prediction for both $q=10$ and $q=20$.
Here $C_o$ is extracted from the peak curvature.
Data from Ref.~[\ref{Billoire 94}].

(\putfig{7b}) Comparison of $2f_{od}$ defined in equation \twof{} with
our prediction without any mixed phase contribution at $q=10$.
Data from Ref.~[\ref{Billoire 94}].

(\putfig{7a}) Comparison of $2f_{od}$ defined and measured in
Ref.~[\ref{Janke 92}]  with
our prediction without any mixed phase contribution at $q=7$.

\vfil\eject
\end